\documentclass[a4paper, 11pt]{article}
\usepackage{amsbsy,amssymb,amsmath,amsfonts}
\usepackage[english]{babel}
\usepackage[latin1]{inputenc}
\usepackage[dvips]{epsfig}
\usepackage[T1]{fontenc}
\usepackage{babel}
\usepackage{xspace}      
 
     \newtheorem{theorem}{Theorem}
     \newtheorem{corollary}{Corollary}
     \newtheorem{lemma}{Lemma}
     \newtheorem{definition}{Definition}
     \newtheorem{example}{Example}
     \newtheorem{proposition}{Proposition}
     \newcommand{\ca}{cellular automaton\xspace}
     \newcommand{\ooo}{\mathcal{O}}
     \newcommand{\cas}{cellular automata\xspace}
     \newcommand{\ie}{\emph{i.e.,}\ }
     \newcommand{\cf}{\emph{cf.}\ }
     \newcommand{\ccc}{\ensuremath{\mathcal{C}}}
     \newcommand{\zed}{{\mathbb Z}}
     \newcommand{\ene}{{\mathbb N}}
     
     \makeatother

\makeatletter
\def\@yproof[#1]{\@proof{ #1}}
\def\@proof#1{\begin{trivlist}\item[]{\em Proof#1.}}
\newenvironment{proof}{\@ifnextchar[{\@yproof}{\@proof{}
}}{{\hfill$\square$}\end{trivlist}}         
     
     \newcommand{\submatNW}[2]{
     \smash{\makebox[0cm][l]{$
     \left.\smash{\underbrace{
       \makebox[0cm][l]{$
         \begin{array}{ccc}
           0&\ldots&0\\
           \vdots&&\vdots\\
           0&\ldots&0
         \end{array}
       $}\rule{12ex}{0pt}
     }_{#1}}\rule{0pt}{5.5ex}\right\}\scriptstyle#2
     $}}\rule{11ex}{0pt}\rule[-5ex]{0pt}{11ex}
     }
\begin{document}

     \title{Number conserving cellular automata: \\ from decidability to
     dynamics}
     
\author{ Bruno Durand, Enrico Formenti\footnote{Laboratoire
d'Informatique de Marseille (LIM), 39 rue Joliot-Curie, 13453
Marseille Cedex 13, France, \texttt{bdurand@cmi.univ-mrs.fr,
eforment@cmi.univ-mrs.fr}}, \\
Zsuzsanna R\'{o}ka\footnote{Laboratoire d'Informatique Th\'eorique et
Appliqu\'ee (LITA), I.U.T.\ de Metz, \^{I}le du Saulcy, 57045 METZ
Cedex 01, France, \texttt{roka@iut.univ-metz.fr}} }

\maketitle

\begin{abstract}

    We compare several definitions for number-conserving cellular
    automata that we prove to be equivalent. A necessary and
    sufficient condition for \cas to be number-conserving is proved. 
    Using this condition, we give a linear-time algorithm to decide
    number-conservation.  
    
    The dynamical behavior of number-conserving \cas is studied and 
    a classification that focuses on chaoticity is given. 
	
\end{abstract}

	
\section{Introduction}

Cellular automata are widely known as formal models for complex
systems ruled by local interactions
\cite{wolfram86,manneville,gutowitz,boccaragoles,bandini}.  Despite of
their apparent simplicity they display a richness of behaviors most of
which are not yet fully understood.  The classification of these
behaviors has a main obstacle: most of the interesting properties
about them are undecidable \cite{kari-rice}.  Moreover, some other
properties are ``dimension sensitive''; for example, surjectivity is
decidable in polynomial time for one-dimensional CA and undecidable
for dimension $\geq 2$ \cite{revers-2D-comp,durjcss93,cervdur}.

We prove that \emph{number-conservation}, a widely used property, can
be decided in linear time (in the size of the CA) in \emph{any}
dimension.  More precisely, it has linear time complexity if we assume
that the access to the transition table costs constant time,
independently of the dimension of the space.  Otherwise the complexity
is $\ooo{(s\log s)}$, where $s$ is the size of the transition table. 
A preliminary work had been carried out for one-dimensional CA on
circular configurations (periodic boundary conditions) by Boccara et
al.\ in \cite{boccara,boccarafuks}.  The decision problem in higher
dimensions and general boundary conditions that we solved in the
present paper is rather surprising and could not be expected from
earlier works.

CA that satisfy number-conservation property are called
number-con\-ser\-ving (NCA); roughly speaking they conserve the number
of bits during their evolution.  This class of CA has been introduced
by \cite{nagel} and is receiving a growing interest in Physics as a
source for models of particles systems ruled by conservations laws of
mass or energy.

Great attention has received the application of NCA to practical
situations such as highway traffic flows.  In this context different
definitions of ``number-conservation'' were given according to the
type of boundary conditions assumed (periodic, finite, infinite)
\cite{nagel,boccara,boccarafuks}.  In Section~\ref{sec.definitions} we
prove that all of them are equivalent.  We also give another
equivalent definition in terms of densities.  Section~\ref{sec.necsuf}
is devoted to the proof of our main result: the linear time decision
algorithm.  In the second part of the paper we study NCA dynamics with
focus on chaotic behavior.  Number-conservation imposes strong
``simplifications'' in the dynamics allowing (in most cases) clear-cut
situations that are not observed in the general case.

We stress that the class of NCA is not only big enough to give
reliable models for car traffic and similar models allowing the study
of traffic jams, critical traffic masses, crossing-roads, car
accidents etc.\ But also they are interesting from a pure computer
science point of view.  For instance Morita et al.\ have given a
reversible two-dimensional NCA capable of universal computation
\cite{morita98,morita99}.
     \section{Definitions}\label{sec.definitions}
We present in this section several definitions of number-conserving
\cas that we prove to be equivalent.  Proofs are given in a
one-dimensional space but can be generalized to any dimension.

Cellular automata are formally defined as quadruples $(d,S,$ $N,f)$. 
The integer $d$ is the \emph{dimension} of the space the \ca will work
on.  $S={\left\{0,1,\ldots,s \right\}}$ is called the set of
{\emph{states}}.  The {\emph{neighborhood}} $N=(x_{1},\ldots x_{v})$
is a $v$-tuple of distinct vectors of $\zed^{d}$.  The $x_i${'}s are
the relative positions of the neighbor cells with respect to the cell,
the new state of which is being computed.  The states of these
neighbors are used to compute the new state of the center cell.  The
\emph{local function} of the cellular automaton $f: S^{v} \mapsto S$
gives the local transition rule.

A {\emph{configuration}} is an application from $\zed^{d}$ to $S$.
The set of configurations is $S^{\zed^{d}}$ on which the 
  \emph{global function} $A$ of the cellular automaton is defined via $f$:
$$\forall c \in S^{\zed^{d}}, \forall i \in \zed^{d},
A{(c)}{(i)} = f{\left(c{(i+x_{1})},\ldots,c{(i+x_{v})}\right)}.$$

We denote by $\ccc_{P}$ the set of (spatial) \emph{periodic
configurations,} (\ie configurations that are periodic in all
dimensions of the space).  We denote by $\pi(c)$ the period of a
periodic configuration (\ie a vector.)  The expression $0\leq k \leq
\pi(c)$, $k\in\zed^d$ means $\forall 1\leq i\leq d,\ 0\leq k_{i} \leq
\pi(c)_{i}$ where $d$ is the dimension of the space and $k_{i}$ is the
$i$-th component of $k$.

     \begin{definition}
             Let \(A\) be a $d$-dimensional \ca. \(A\) is said to be
             \emph{periodic-number-conserving (PNC)} iff
             $$\displaystyle  \forall c\in \ccc_{P},
             \ \sum_{0\leq k \leq \pi(c)}c(k)= 
             \sum_{0\leq k \leq \pi(c)}A(c)(k).$$
     \end{definition}

We denote by $\ccc_{F}$ the set of \emph{finite configurations,} (\ie
those configurations that are almost everywhere equal to zero ---
non-zero in a finite number of cells). 
     \begin{definition}
             Let \( A \) be a $d$-dimensional \ca.  \( A \) is said to be
             \emph{finite-number-conserving (FNC)} iff
             $$\displaystyle  \forall c\in \ccc_{F},
             \ \sum_{i\in \zed^d }c(i)=\sum_{i\in \zed^d}A(c)(i).$$
     \end{definition}

     We denote by $\ccc$ the set of \emph{all configurations.}

     \begin{definition}
	Let \( A \) be a $d$-dimensional \ca.  A \emph{window} is a
	hypercube of $\zed^d$ centered in $\mathbf{0}$ thus determined
	by its size.  Consider the sequence of windows
	$\{F_n\}$ of size $2n+1$ and denote by $\mu_n(c)$ the sum of
	states in $F_n$ of a configuration $c\in\ccc$.  $A$ is said to
	be \emph{number-conserving (NC)} iff $$\forall c\in\ccc,
	\displaystyle
     	\lim_{n\to \infty}\frac{\mu_n(c)}{\mu_n(A(c))}=1.$$
     \end{definition}
     
     Remark that all these definitions imply that $f(0,0,\ldots,0)=0$,
     \ie $0$ is a \emph{quiescent} state. The configuration with all
     quiescent states is denoted $uz$.

     We are going to prove that all previous 
     definitions are equivalent. The proof is given in dimension~1; 
     the generalization to higher dimensions is straightforward.    
     \begin{proposition}\label{prop:equivalence}
             A cellular automaton $A$ is FNC if and only if it is PNC.
     \end{proposition}
     \begin{proof}
     We first prove that if $A$ is PNC then it is FNC. We show that if
     $A$ is not FNC then it is not PNC either. 

     Let \( A \) be not FNC, then $$\displaystyle \exists
     \widetilde{c}\in \ccc_{F}, \exists \alpha,\beta\in\ene, \
     \alpha=\sum _{i \in \zed}\widetilde{c}(i)\neq \sum _{i \in \zed
     }A(\widetilde{c})(i)=\beta.$$

     Let \( k \) be the length of the non-zero part of \(
     \widetilde{c} \) and \( l \) the radius of the neighborhood of
     $A$.  Then, the length of the non-zero part of \(
     A(\widetilde{c}) \) is at most \( k+2l \) (see
     Figure~\ref{fig:per}).

     Let \( \widehat{c} \) be the periodic configuration constructed from
     \( \widetilde{c} \) in the following way.  We add \( l \) zeros at
     both sides of the non-zero part of \( \widetilde{c} \) and we
     construct a periodic configuration with the segment obtained (such a
     configuration has a period \( p=k+2l\), see Figure~\ref{fig:per}). 

     \begin{figure*}[hbt]
            \centerline{\psfig{file=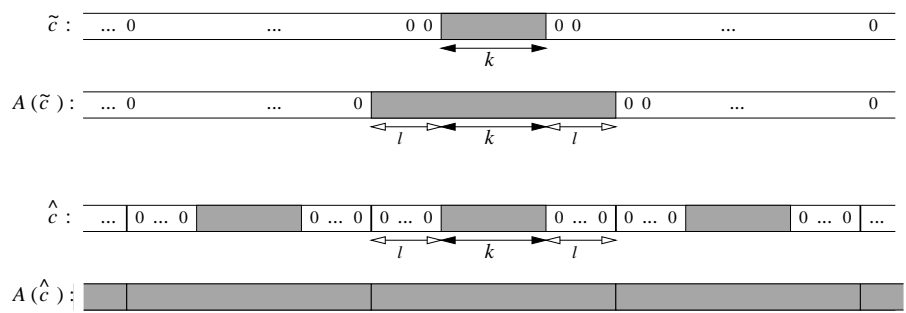,width=12cm}}
             \caption{Construction of the periodic configuration}
             \label{fig:per}
     \end{figure*}

     We have
     $$\displaystyle \sum _{0\leq i\leq p}\widehat{c}(i)= \alpha \neq
     \beta =\sum _{0\leq i \leq p}A(\widehat{c})(i),$$ hence \( A \) is not
     PNC.

     \medskip We now prove that if $A$ is FNC then it is PNC. We show
	that if $A$ is not PNC then it is not FNC either.

     Let \( l \) be the radius of the neighborhood.  As \( A \) is not
     PNC, $$\displaystyle \exists p, \alpha, \beta \in \ene, \exists
     \widehat{c}\in \ccc_{P}, \ \sum _{0\leq i \leq p}\widehat{c}(i) =
     \alpha \neq \beta = \sum _{0\leq i \leq p}A(\widehat{c})(i),$$
     where \( p \) is the period of \( \widehat{c} \). \vspace{3mm}

     We denote by 
	\begin{description}
		\item \hspace{1cm} $\widehat{c}_{p}$ the $p$-length
             		segment of $\widehat{c}$ centered at the origin;
             	\item \hspace{1cm} $\widehat{c}_{p\textnormal{Suffix}_{l}}$
             	 	the $l$-length \textnormal{suffix} of 
			$\widehat{c}_{p}$;
             	\item \hspace{1cm} $\widehat{c}_{p\textnormal{Prefix}_{l}}$ 
			the $l$-length \textnormal{prefix} of
			$\widehat{c}_{p}$.
          \end{description}  
 
     We construct a finite configuration $\widetilde{c}\in \ccc_{F_{A}}$
     as shown in Figure~\ref{fig:finite}:
     we repeat $\widehat{c}_{p}$ $x$ times and we complete the
     segment obtained in this way with 
     $\widehat{c}_{p\textnormal{Prefix}_{l}}$ at the right-hand-side and with
     $\widehat{c}_{p\textnormal{Suffix}_{l}}$ at the left-hand-side of
	the segment.  

     \begin{figure*}[hbt]
             \centerline{\psfig{file=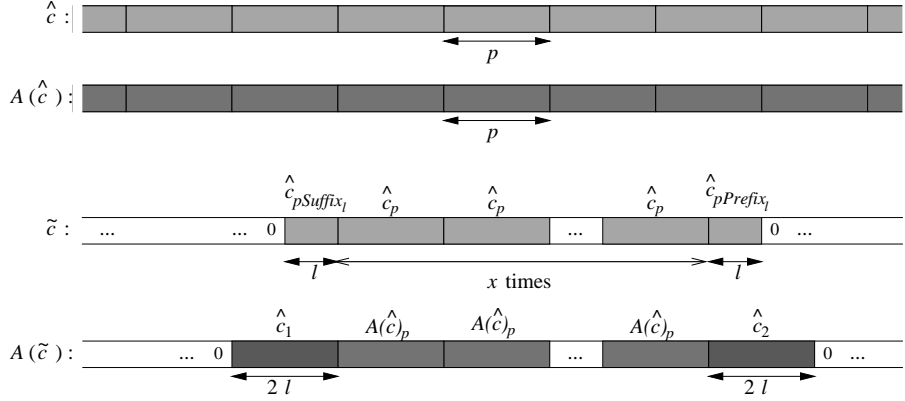,width=12cm}}
             \caption{Construction of the finite configuration}
             \label{fig:finite}
     \end{figure*}

        Then, $$ \displaystyle \sum _{i\in\zed}
        \widetilde{c}(i)=\underbrace{x\sum_{i=1}^{p}
        \widehat{c}_{p}(i)}_{x\alpha }+\underbrace{\sum_{i=1}^{l}
        \widehat{c}_{p\textnormal{Prefix}_{l}}(i)}_{\leq ls}+
        \underbrace{\sum_{i=1}^{l}
        \widehat{c}_{p\textnormal{Suffix}_{l}}(i)}_{\leq ls}$$  
        and
        \begin{eqnarray*}
         \sum _{i\in\zed}
        A(\widetilde{c})(i) & = & \underbrace{x\sum_{i=1}^{p}
        A(\widehat{c}_{p})(i)}_{x\beta }+\underbrace{\sum_{i=1}^{l}
        A(\widehat{c}_{p\textnormal{Prefix}_{l}})(i)}_{\leq 2ls} + 
	\underbrace{\sum_{i=1}^{l}
        A(\widehat{c}_{p\textnormal{Suffix}_{l}})(i)}_{\leq 2ls}. 
        \end{eqnarray*}

	We obtain the following inequalities: 
	$$x\alpha\leq\sum _{i\in\zed}\widetilde{c}(i)\leq x\alpha+2ls 
	\textnormal{ and }$$  
	$$x\beta \leq \sum_{i\in\zed} A(\widetilde{c})(i)\leq x\beta +4ls.$$

     Suppose that \( \alpha <\beta \).  If we choose an integer $x$ such
     that \( x>\frac{2ls}{\beta -\alpha } \), we are in a contradiction with
     the previous inequalities.  

     Similarly, if \( \alpha >\beta \) then the contradiction is
     obtained by choosing \( x>\frac{4ls}{\alpha -\beta } \).  

	Hence, if \( A \) is not PNC, it is not FNC either.
     \end{proof}

     \begin{proposition}
     A cellular automaton $A$ is NC if and only if it is FNC.
     \end{proposition}
     \begin{proof}
We first show that if $A$ is not FNC then it is not NC either.  Let
$A$ be a cellular automaton with a neighborhood of radius $l$.  If $A$
is not FNC then there exist an integer $\gamma$ and a configuration
$c$ in $\ccc_{F}$ such that
$$\sum_{i\in\zed}c(i)=\sum_{i=-\gamma}^{\gamma}c(i) \not
=\sum_{i=-\gamma -l}^{\gamma +l}A(c)(i)=\sum_{i\in\zed}A(c)(i).$$ Let
$\{F_n\}$ be a sequence of windows of size tending to the infinity,
centered on the non-quiescent part of $c$.  As the non-quiescent part
of $c$ is finite, there exists $k$ such that, for all $n\geq k$,
$$\mu_n(c)=\sum_{i=-\gamma }^{\gamma}c(i) \not =
	\sum_{i=-\gamma -l }^{\gamma +l}A(c)(i)=\mu_n(A(c)).$$ 
Hence, $\displaystyle
\lim_{n\to=\infty}\frac{\mu_n(c)}{\mu_n(A(c))}\not =1$, so $A$ is not
NC.\vspace{5mm}
	
Now assume that $A$ is not NC. Then there exists a configuration $c$
such that $\displaystyle \frac{\mu_n(c)}{\mu_n(A(c))}$ does not
converge to 1.  Now let us call frontier of the window the stripe of
width the diameter of the neighborhood at the window border.  Let us
now count $\xi_{n}(c)$ the number of non-zero cells of $c$ in this
frontier.

First case: $\forall a, \exists n,\ \mu_{n}<an$.  Thus there exist
consecutive subsequences of unbounded size in which $\mu_{n}(c)$ is
constant.  Let us consider such a subsequence of length greater than
the neighborhood.  The frontier of the last window is thus full of
zeros.  Hence we can ``cut off'' the internal part, surround it by 0's
and thus obtain a finite configuration.  As $A$ is not NC, it is not
FNC either.

Second case: $\exists a, \forall n,\ \mu_{n}>an$.  The idea is to
``cut off'' the windows to form periodic configurations.  We can bound
the perturbation on $\mu_{n}(A(x)$ by the number of non-zero cells in
the frontier multiplied by the maximal state (a constant).  We then
remark that with the hypothesis above, $\xi_{n}(c)=o(\mu_{n}(c))$. 
Thus for $n$ sufficiently big, the unbalance of the ratio
$\frac{\mu_n(c)}{\mu_n(A(c))}$ is transmitted to the periodic 
configuration obtained. Therefore $A$ is not PNC and so not FNC.
     \end{proof}

\section{A necessary and sufficient condition}\label{sec.necsuf}

We present below a necessary and sufficient condition for
number-conservation.  Our condition is rather complicated to
construct.  For the sake of simplicity, we first present it in
dimension~2 with a square neighborhood $2\times 2$.  Then we give the
condition for  neighborhood $n\times m$ (proof sketched).
     
     \begin{proposition}\label{example}
             Let $A=(2,N,Q,f)$ be a 2D cellular automaton with $N=\left( 
             \begin{array}{ll}
                     (0,0) & (0,1)\\
                     (0,-1) & (1,-1)
             \end{array}\right)$ and thus  $f:Q^4\to Q$.
     $A$ is number-conserving if and only if
     {\small \begin{eqnarray*}
     f\left( 
             \begin{array}{cc}
                     a & b\\
                     c & d
             \end{array}\right) 
     & = a & 
             +\left[ f\left( 
                     \begin{array}{cc}
                             0 & 0\\
                             0 & c
                     \end{array}\right) 
             -f\left( 
                     \begin{array}{cc}
                             0 & 0\\
                             0 & a
                     \end{array}\right) \right] \\
     &   &
             +\left[ f\left( 
                     \begin{array}{cc}
                             0 & 0\\
                             0 & b
                     \end{array}\right) 
             -f\left( 
                     \begin{array}{cc}
                             0 & 0\\
                             0 & d
                     \end{array}\right) \right] \\
     &   & 
             +\left[ f\left( 
                     \begin{array}{cc}
                             0 & 0\\
                             c & d
                     \end{array}\right) 
             -f\left( 
                     \begin{array}{cc}
                             0 & 0\\
                             a & b
                     \end{array}\right) \right] \\
     &   &
             +\left[ f\left( 
                     \begin{array}{cc}
                             0 & b\\
                             0 & d
                     \end{array}\right) 
             -f\left( 
                     \begin{array}{cc}
                             0 & a\\
                             0 & c
                     \end{array}\right) \right].
     \end{eqnarray*}}
     \end{proposition}
     \begin{proof}
     We first show that the condition is necessary.  Assume that $A$ is NC,
     let us denote by $\natural c$ the sum of the states in a
     finite configuration $c$ of $A$.  As $A$ is in particular FNC (see 
     Prop.~\ref{prop:equivalence}) 
             {\small \[ \natural A\left( \begin{array}{cccccc}
             ... & ... & ... & ... & ... & ...\\
             ... & 0 & 0 & 0 & 0 & ...\\
             ... & 0 & a & b & 0 & ...\\
             ... & 0 & c & d & 0 & ...\\
             ... & 0 & 0 & 0 & 0 & ...\\
             ... & ... & ... & ... & ... & ...
             \end{array}\right)
             =a+b+c+d. \]}
     \noindent Moreover, 
     {\small \begin{eqnarray}
     a+b+c+d & = &  f\left( 
                     \begin{array}{cc}
                             0 & 0\\
                             0 & a
                     \end{array}\right) 
           	+f\left( 
                     \begin{array}{cc}
                             0 & 0\\
                             b & 0
                     \end{array}\right) 
        	+f\left( 
                     \begin{array}{cc}
                             0 & c\\
                             0 & 0
                     \end{array}\right) 
		+f\left( 
                     \begin{array}{cc}
                             d & 0\\
                             0 & 0
                     \end{array}\right) \nonumber \\
 & &
         	+f\left( 
                     \begin{array}{cc}
                             0 & 0\\
                             a & b
                     \end{array}\right) 
      		+f\left( 
                     \begin{array}{cc}
                             c & d\\
                             0 & 0
                     \end{array}\right) +f\left( 
                     \begin{array}{cc}
                             0 & a\\
                             0 & c
                     \end{array}\right) 
              +f\left( 
                     \begin{array}{cc}
                             b & 0\\
                             d & 0
                     \end{array}\right)\nonumber \\
& & 
           	+f\left( 
                     \begin{array}{cc}
                             a & b\\
                             c & d
                     \end{array}\right)\label{abcd} \end{eqnarray}}
     \noindent Replacing $a$ by 0 yields:
     \noindent{\small \begin{eqnarray}
     b+c+d  & = &   f\left( 
                     \begin{array}{cc}
                             0 & 0\\
                             b & 0
                     \end{array}\right) 
             	+ f\left( 
                     \begin{array}{cc}
                             0 & 0\\
                             0 & b
                     \end{array}\right) 
             	+ f\left( 
                     \begin{array}{cc}
                             0 & c\\
                             0 & 0
                     \end{array}\right)
		+ f\left( 
                     \begin{array}{cc}
                             0 & 0\\
                             0 & c
                     \end{array}\right)   \nonumber \\
              &  &
             	+ f\left( 
                     \begin{array}{cc}
                             d & 0\\
                             0 & 0
                     \end{array}\right)  
		+f\left( 
                     \begin{array}{cc}
                             c & d\\
                             0 & 0
                     \end{array}\right) 
		+f\left( 
                     \begin{array}{cc}
                             b & 0\\
                             d & 0
                     \end{array}\right) 
             	+f\left( 
                     \begin{array}{cc}       
                             0 & b\\
                             c & d
                     \end{array}\right) \label{bcd}
     \end{eqnarray}}
     \noindent Subtracting (\ref{bcd}) from (\ref{abcd}) we obtain:
     {\small \begin{eqnarray}
     f\left( 
             \begin{array}{cc}
                     a & b\\
                     c & d
             \end{array}\right) 
     & = a &  +f\left( 
                     \begin{array}{cc}
                             0 & 0\\
                             0 & b
                     \end{array}\right) 
             +f\left( 
                     \begin{array}{cc}
                             0 & 0\\
                             0 & c
                     \end{array}\right) 
             +f\left( 
                     \begin{array}{cc}
                             0 & b\\
                             c & d
                     \end{array}\right) \nonumber \\
     &  & 
             - f\left( 
                     \begin{array}{cc}
                             0 & 0\\
                             0 & a
                     \end{array}\right) 
             -f\left( 
                     \begin{array}{cc}
                             0 & 0\\
                             a & b
                     \end{array}\right)      
             -f\left( \begin{array}{cc}
                             0 & a\\
                             0 & c
                     \end{array}\right) \label{fabcd}
     \end{eqnarray}}
     \noindent Setting \( b=0 \) in (\ref{bcd}), we have also:
     {\small\begin{eqnarray}
     c+d & = & 
	f\left( 
                     \begin{array}{cc}
                             0 & 0\\
                             0 & c
                     \end{array}\right) 
	+f\left( 
                     \begin{array}{cc}
                             0 & c\\ 
                             0 & 0
                     \end{array}\right) 
	+f\left( 
                     \begin{array}{cc}
                             0 & 0\\
                             d & 0
                     \end{array}\right)   
	+f\left( 
                     \begin{array}{cc}
                             d & 0\\
                             0 & 0
                     \end{array}\right)\nonumber  \\
              & &
	+f\left( 
                     \begin{array}{cc}
                             0 & 0\\
                             c & d
                     \end{array}\right)
	+f\left( 
                     \begin{array}{cc}
                             c & d\\
                             0 & 0
                     \end{array}\right) \label{cd} 
     \end{eqnarray}}
     \noindent Subtracting (\ref{cd})  from (\ref{bcd}) we obtain:
     {\small \begin{eqnarray}
     f\left( 
             \begin{array}{cc}
                     0 & b\\
                     c & d
             \end{array}\right) 
     & = b &
             + f\left( 
                     \begin{array}{cc}
                             0 & 0\\
                             d & 0
                     \end{array}\right) 
             +f\left( 
                     \begin{array}{cc}
                             0 & 0\\
                             c & d
                     \end{array}\right) 
	-f\left( 
                     \begin{array}{cc}
                             0 & 0\\
                             0 & b
                     \end{array}\right) \nonumber \\
     & &
             -f\left( 
                     \begin{array}{cc}
                             0 & 0\\
                             b & 0
                     \end{array}\right) 
             -f\left( 
                     \begin{array}{cc}
                             b & 0\\
                             d & 0
                     \end{array}\right) \label{fbcd}
     \end{eqnarray}}
     \noindent We replace this term in (\ref{fabcd}):
     {\small \begin{eqnarray}
     f\left( 
             \begin{array}{cc}
                     a & b\\
                     c & d
             \end{array}\right) 
     & = & 
             a + b  
		-f\left( 
                     \begin{array}{cc}
                             b & 0\\
                             d & 0
                     \end{array}\right) 
		+ f\left( 
                     \begin{array}{cc}
                             0 & 0\\
                             c & d
                     \end{array}\right) 
	-f\left( 
                     \begin{array}{cc}
                             0 & 0\\
                             a & b
                     \end{array}\right)
	+ f\left( 
                     \begin{array}{cc}
                             0 & 0\\
                             0 & c
                     \end{array}\right) \nonumber \\
	& & 
             	-f\left( 
                     \begin{array}{cc}
                             0 & 0\\
                             0 & a
                     \end{array}\right) 
             	+f\left( 
                     \begin{array}{cc}
                             0 & 0\\
                             d & 0
                     \end{array}\right)
             	-f\left( 
                     \begin{array}{cc}
                             0 & 0\\ 
                             b & 0
                     \end{array}\right)
             	-f\left( 
                     \begin{array}{cc}
                             0 & a\\
                             0 & c
                     \end{array}\right) \label{fabcd2}
     \end{eqnarray}}
     \noindent If $A$ is FNC then 
     {\small \[\natural A\left( \begin{array}{cccccc}
             ... & ... & ... & ... & ... & \\
             ... & 0 & 0 & 0 & ... & \\
             ... & 0 & b & 0 & ... & \\
             ... & 0 & d & 0 & ... & \\
             ... & 0 & 0 & 0 & ... & \\
             ... & ... & ... & ... & ...
     \end{array}\right) = b+d. \]}
      \noindent After similar calculi as above with the substitution
      of $b$ by 0, we obtain  
     {\small \begin{eqnarray}
     f\left( 
             \begin{array}{cc}
                     b & 0\\
                     d & 0
             \end{array}\right) 
     & = & 
             b+f\left( 
                     \begin{array}{cc}
                             0 & 0\\
                             0 & d
                     \end{array}\right) 
             +f\left( 
                     \begin{array}{cc}
                             0 & 0\\ 
                             d & 0
                     \end{array}\right) -f\left( 
                     \begin{array}{cc}
                             0 & 0\\
                             0 & b
                     \end{array}\right)-f\left( 
                     \begin{array}{cc}
                             0 & 0\\
                             b & 0
                     \end{array}\right)\nonumber \\
     &   &               
             -f\left( 
                     \begin{array}{cc}
                             0 & b\\
                             0 & d
                     \end{array}\right)\label{fbd}
     \end{eqnarray}}
     \noindent By replacing this term in (\ref{fabcd2}), we obtain
exactly the
        formula of our Proposition. 

     Let us prove now that this condition implies PNC. Then by
     Prop.~\ref{prop:equivalence}, the theorem is proved.  Consider a
     periodic configuration.  Now sum this condition on the period; the
     terms $[f(\ldots)-f(\ldots)]$ simplify and we get exactly the 
     condition for PNC hence for NC.
     \end{proof}
     We now give the condition for more general neighborhood in
     dimension~2.  We assert that an analogous formula can be obtained
     for higher dimension.

     \begin{theorem}\label{toto}
             Let $A=(2,N,Q,f)$ be a 2D cellular automaton with 
     $$N=\left( \begin{array}{llll}
     (0,0)&(0,1)&\ldots&(0,n-1)\\
     (1,0)&(1,1)&\ldots&(1,n-1)\\
     \vdots& \vdots& & \vdots\\
     (m-1,0)&(m-1,1)&\ldots&(m-1,n-1)\end{array}\right)$$
     and thus $f:Q^{m\times n}\rightarrow Q$.  $A$ is
     number-conserving if and only if
     $$ f\left( \begin{array}{cccc}
     x_{1,1} & x_{1,2} & \ldots & x_{1,n}  \\
     x_{2,1} & x_{2,2} & \ldots & x_{2,n} \\
     \vdots & \vdots &  & \vdots \\
     x_{m,1} & x_{m,2} & \ldots & x_{m,n} 
     \end{array}\right) = x_{1,1}$$

     \newcommand{\submatSE}{
       \begin{array}{ccc}
         x_{1,2}&\ldots&x_{1,n-j-1}\\
         x_{2,2}&\ldots&x_{2,n-j-1}\\
         \vdots&\vdots&\vdots\\
         x_{m-i,2}&\ldots&x_{m-i,n-j-1}
       \end{array}
     }

    {\small  \[
     +\sum ^{m-1}_{i=0}\sum ^{n-2}_{j=0} f\left(
       \begin{array}{cc}
         \submatNW{j+1}{i} & 0 \\
         0 & \submatSE
       \end{array}
     \right)
     \]}

     \newcommand{\submat}{
       \begin{array}{ccc}
         x_{2,1}&\ldots&x_{2,n-j}\\
         x_{3,1}&\ldots&x_{3,n-j}\\
         \vdots&\vdots&\vdots\\
         x_{m-i-1,1}&\ldots&x_{m-i-1,n-j}
       \end{array}
     }

     {\small \[
     +\sum ^{m-2}_{i=0}\sum ^{n-1}_{j=0} f\left(
       \begin{array}{cc}
         \submatNW{j}{i+1} & 0 \\
         0 & \submat
       \end{array}
     \right)
     \]}

     \newcommand{\submata}{
       \begin{array}{ccc}
         x_{1,1}&\ldots&x_{1,n-j}\\
         x_{2,1}&\ldots&x_{2,n-j}\\
         \vdots&\vdots&\vdots\\
         x_{m-i,1}&\ldots&x_{m-i,n-j}
       \end{array}
     }

    {\small  \[
     -\sum ^{m-1}_{i=0}\sum ^{n-1}_{j=0 (i+j>0)} f\left(
       \begin{array}{cc}
         \submatNW{j}{i} & 0 \\
         0 & \submata
       \end{array}
     \right)
     \]}

     \newcommand{\submatb}{
       \begin{array}{ccc}
         x_{2,2}&\ldots&x_{2,n-j-1}\\
         x_{3,2}&\ldots&x_{3,n-j-1}\\
         \vdots&\vdots&\vdots\\
         x_{m-i-1,2}&\ldots&x_{m-i-1,n-j-1}
       \end{array}
     }

     {\small \[
     -\sum ^{m-2}_{i=0}\sum ^{n-2}_{j=0} f\left(
       \begin{array}{cc}
         \submatNW{j+1}{i+1} & 0 \\
         0 & \submatb
       \end{array}
     \right)
     \]}

     \end{theorem}

     \begin{proof}
     We give only the sketch of the proof.
     In order to show that the condition is necessary, we start from the
     configuration   
     {\small $$ \begin{array}{cccccccc}
            & \vdots & \vdots  & \vdots  &        & \vdots  & \vdots & \\
     \ldots & 0      & 0       & 0       &        & 0       & 0      &
\ldots \\
     \ldots & 0      & x_{1,1} & x_{1,2} & \ldots & x_{1,n} & 0      &
\ldots \\
     \ldots & 0      & x_{2,1} & x_{2,2} & \ldots & x_{2,n} & 0      &
\ldots \\
     \vdots & 0      & \vdots  & \vdots  &        & \vdots  & 0      &
\vdots \\
     \ldots & 0      & x_{m,1} & x_{m,2} & \ldots & x_{m,n} & 0      &
\ldots \\
     \ldots & 0      & 0       & 0       &        & 0       & 0      &
\ldots \\
            & \vdots & \vdots  & \vdots  &        & \vdots  & \vdots & 
     \end{array}$$}
     As $A$ is FNC, we have $x_{1,1}+x_{1,2}+\ldots + x_{m,n} = $ sum of
     states obtained in each cell after the application of $A$, that is the
     sum of all $f(\ldots)$; we substitute in this equation $x_{1,1}$ by 0,
     then $x_{1,2}$ by 0, and so on, up to $x_{1,n}$.  Then we obtain
     an analogous formula to~(\ref{fabcd2})
     {\small $$ f\left( \begin{array}{cccc}
     x_{1,1} & x_{1,2} & \ldots & x_{1,n}  \\
     x_{2,1} & x_{2,2} & \ldots & x_{2,n} \\
     \vdots & \vdots &  & \vdots \\
     x_{m,1} & x_{m,2} & \ldots & x_{m,n} 
     \end{array}\right) = x_{1,1} + x_{1,2} + \ldots + x_{1,n} $$}
     {\small $$ - 
                     f\left( \begin{array}{ccccc}
                     x_{1,2} & x_{1,3} & \ldots & x_{1,n} & 0      \\
                     x_{2,2} & x_{2,3} & \ldots & x_{2,n} & 0      \\
                     \vdots  &         & \vdots & \vdots  & \vdots \\
                     x_{m,2} & x_{1,2} & \ldots & x_{m,n} & 0
                     \end{array}\right) + \ldots$$}
     Then, in order to obtain the final formula, starting from the configuration
     {\small $$ \begin{array}{cccccccc}
            & \vdots & \vdots  & \vdots  &        & \vdots  & \vdots &  
     \\
     \ldots & 0      & 0       & 0       &        & 0       & 0      &
\ldots \\
     \ldots & 0      & x_{1,2} & x_{1,3} & \ldots & x_{1,n} & 0      &
\ldots \\
     \ldots & 0      & x_{2,2} & x_{2,3} & \ldots & x_{2,n} & 0      &
\ldots \\
     \vdots & 0      & \vdots  & \vdots  &        & \vdots  & 0      &
\vdots \\
     \ldots & 0      & x_{m,2} & x_{m,3} & \ldots & x_{m,n} & 0      &
\ldots \\
     \ldots & 0      & 0       & 0       &        & 0       & 0      &
\ldots \\
            & \vdots & \vdots  & \vdots  &        & \vdots  & \vdots & 
     \end{array}$$}
     we establish the equation for 
     {\small $$ f\left( \begin{array}{ccccc}
     x_{1,2} & x_{1,3} & \ldots & x_{1,n} & 0  \\
     x_{2,2} & x_{2,3} & \ldots & x_{2,n} & 0\\
     \vdots & \vdots &  & \vdots & \vdots \\
     x_{m,2} & x_{m,3} & \ldots & x_{m,n} & 0
     \end{array}\right).$$}
     Then, we substitute $x_{1,2}$ by 0, then $x_{1,3}$, and so on, up to
     $x_{1,n}$. 

     The condition is sufficient: for periodic configurations, when
     calculating the sum of states on the period, every $f$-term is
     present two times. If it is present in the sum of $f$-terms with
     $x_{1,1}$ (resp. $x_{2,2}$) as first non-zero element, then it is
     present in the sum of $f$-terms with $x_{1,2}$ (resp. $x_{2,1}$)
     as first non-zero element. We so have as sum of states on a
     period $\sum_{i=1}^{m}\sum_{j=1}^{n}x_{i,j}$. Hence this
     condition implies PNC and by Prop.~\ref{prop:equivalence} the
     theorem is proved.
\end{proof}

     \begin{theorem}
         The property ``number-conserving'' is decidable in 
         $\ooo(s\log s)$ for cellular automata of any dimension, where 
         $s$ denotes the size of the transition table.
     \end{theorem}

     \begin{proof}
         Remark that the number of terms of our condition in
Th.~\ref{toto} 
         is proportional to the number of neighbors. Thus  the set of
         conditions to be checked is of size proportional to the size of 
         the local transition rule of the cellular automata, which is
upper 
         bounded by the size of the cellular automaton in the chosen 
         representation.  Access to the table costs at most 
         $\ooo(\log s)$.
     \end{proof}

     Remark that if \cas are given by a program computing their rule and 
     not by the rule exhaustively, then our theorem does not hold. This 
     drawback is often present in \cas theory. But if we consider that 
     access to the transition table is performed in constant time, 
     then the algorithm is linear.
%
%
%

\section{Dynamics}

\newcommand{\uz}{\underline{0}}
\newcommand{\zu}{\set{0,1}}
\newcommand{\len}[1]{\ell\left(#1\right)}
\newcommand{\CC}{\mathcal{C}}
\newcommand{\FF}{\mathcal{F}}
\newcommand{\set}[1]{\left\{#1\right\}}

NCA are often used for the simulation of real phenomena
such as highway traffic flow, fluid dynamics and so on.
Therefore, once we have established that our model
is a NCA (by using the decision procedures given in the
first part of the paper, for example), it is very important
to forecast as much as possible its dynamics and, whenever
possible, correlate experimental evidence (such as pattern growth)
with other interesting theoretical properties. This is the matter
of this second part of the paper.

\subsection{Definitions and background}
A \emph{dynamical system} $(X,F)$ consists of a compact metric space
$X$ and a continuous self-map $F$.  Denote $d$ the metric on $X$ and
$F^n$ the $n$-fold composition of $F$ with itself.

We first recall some classical definitions.  A point $x\in X$ is an
\emph{equicontinuity} point for $F$ if for any $\epsilon>0$ there
exists $\delta>0$ such that for any $y\in X$, $d(x,y)\leq\delta$
implies that $\forall t\in \ene$, $d(F^t(x),F^t(y))<\epsilon$.  If $X$ is
perfect and all of its points is an equicontinuity point for $F$ then
$F$ is \emph{equicontinuous}.  A dynamical system $(X,F)$ is
\emph{sensitive to initial conditions} if there exists $\epsilon>0$
such that for any $x\in X$ and any $\delta>0$, $0<d(x,y)<\delta$
implies that there exists $t\in \ene$ such that
$d(F^t(x),F^t(y))\geq\epsilon$.  We recall that in a perfect space any
system with no equicontinuity points is sensitive to initial
conditions and vice-versa.  $(X,F)$ is \emph{expansive} if there
exists $\epsilon>0$ such that for any $x,y\in X$, $d(x,y)>0$ implies
that $\exists t\in \ene$ such that $d(F^t(x),F^t(y))\geq\epsilon$.  If
for any pair of non-empty open set $U,V$ there exists $n\in \ene$ such
that $F^n(U)\cap V\ne\emptyset$ then $F$ is \emph{transitive}.  A
point $x$ is \emph{periodic} if there exists $n\in \ene$ such that
$F^n(x)=x$.  $F$ is \emph{regular} if periodic points are dense in
$X$.

Properties like sensitivity, expansivity, transitivity and regularity
are often used as basic components of \emph{chaotic behavior.} For
example, Devaney, in a popular book \cite{devaney89}, defines as
{chaotic} all systems that are transitive, regular and sensitive to
initial conditions.

We are interested in the study of dynamical systems in symbolic
spaces.  $S^{Z^k}$ is endowed with the product topology (also called
Cantor topology) of a countable product of discrete space.  For the
sake of simplicity, we will study \cas in dimension $k=1$, the
generalization of results to higher dimensions is straightforward for
most of the results; Propositions \ref{prop.bltisseur} to
\ref{prop.c2.k2} require deeper adaptations although no new idea is
needed.

The metric $d$ on $S^{\zed}$ defined as
\[
    \forall x,y\in S^{\zed},\;d(x,y)=2^{-n}
\]

where $n=\inf\set{i\geq 0,x(i)\ne y(i)\;\textrm{or}\;x(-i)\ne y(-i)}$
induces the product topology on $S^{\zed}$. Denote by $S^{\star}$ the
set of finite words on $S$. For $w\in S^{\star}$, $\len{w}$ denotes the
length of $w$.
For $t\in \zed, w\in S^{\star}$, the sets 
\[
  [w]_t=\set{x\in S^{\zed}, x(t)=w(0),\ldots,x(t+\len{w})=w(\len{w})}
\]
are called \emph{cylinders} and form a basis for the product topology
on $S^{\zed}$.  The \emph{shift map} $\sigma$ is often used as a
paradigmatic example of chaotic symbolic system.  It is defined as
$\forall c\in S^{\zed},\forall i\in \zed,\ \sigma(c)(i)=c(i+1)$. 
Cellular automata are exactly those continuous maps from $S^{\zed}$ to
$S^{\zed}$ that commute with the shift (\ie $F\circ\sigma=\sigma\circ
F$)~\cite{hedlund69}.
\subsection{Classification of dynamical behavior}
 \label{subsec.dynamics}
  In spite of their very simple definition \cas may have complex
  dynamical behavior (see for example \cite{wolfram86}).  The
  classification of these dynamical behaviors is known as the
  \emph{classification problem} is one of the oldest, hardest and for
  this reason fascinating open problem in \cas theory.  During the
  years many authors have proposed partial solutions (because of their
  undecidability).  Each classification focuses on a particular aspect
  of \cas theory.  For example, if one is interested in simulations of
  real life phenomena like car traffic (e.g.\ using number-conserving
  \cas that we call NCA for short), the study of dynamics on finite
  patterns deserves a particular importance.  In this context a
  well-known classification of \cas with quiescent state is given in
  \cite{braga93,braga95}.  In these papers \cas are classified
  according to their pattern divergence.  Three classes of behavior
  are defined:
  \begin{itemize}
    \item[$\CC_1$:] $\forall x\in\ccc_F\;\lim_{t\to\infty}\len{F^t(x)}=0$
    \item[$\CC_2$:] $\forall x\in\ccc_F\;\sup_{t\in \ene}\len{F^t(x)}<\infty$
    \item[$\CC_3$:] $\exists x\in\ccc_F\;\sup_{t\in \ene}\len{F^t(x)}=\infty$
  \end{itemize}
  People familiar with the study of complex analytic transformations
  will immediately remark the strong analogies with Julia and Fatou
  sets of complex transformations \cite{blanchard}.

  This classification is especially pertinent  for NCA since $0$ is 
  necessarily a quiescent state.

  \smallskip
  In the study of chaotic behavior one often finds that complex
  behavior is generated from ``simple'' local interactions. 
  Again, using the metaphor of car traffic, traffic-jams and
  their consequences are generated by a ``bad'' interaction of
  a relatively small number of cars. In \cite{kurka97}, K\r{u}rka
  proposed a classification based on local behavior of \cas and 
  increasing degrees of chaos. K\r{u}rka proposed the following
  classes:
  \begin{itemize}
    \item[$K_1$:]
        equicontinuous \cas;
    \item[$K_2$:]
       \cas with equicontinuity points but not equicontinuous;
    \item[$K_3$:]
       \cas sensitive to initial conditions but not expansive;
    \item[$K_4$:]
        expansive \cas.
  \end{itemize}
  The following result is adapted from Knudsen
\cite{knudsen94a:finite}. 
  By using Proposition \ref{prop.knud} one can express the properties
  defining (most of) K\r{u}rka classes in term of behavior on finite
  patterns.
  \begin{proposition}
    \label{prop.knud}
    $F$ is transitive (resp. sensible to initial conditions, 
	e\-qui\-con\-ti\-nuous) 
    w.r.t.   finite configurations if and only if $F$ is 
    transitive (resp. sensible to initial conditions, equicontinuous).
  \end{proposition}
  The same result holds if in Proposition \ref{prop.knud} we consider
the 
  set of spatial periodic configurations instead of finite configurations.
  In the rest of the section, we study the relations between 
  dynamical behavior of NCA and the classifications
  of K\r{u}rka and Cattaneo.
  
  Given $x\in \ccc_F$, let $$m(x)=\min\set{i\in \zed,\,x(i)\ne 0},\ 
  M(x)=\max\set{i\in \zed,\,x(i)\ne 0}\ {\rm and}$$
  $$s(x)=\sum^{M(x)}_{m(x)} x(i).$$
  Remark that $\forall x\in \ccc_F,\;m(\sigma^k(x))=m(x)-k$ (the same
  holds if we put $M$ in place of $m$).
  
\begin{proposition}
  The are no expansive NCA.
  \label{prop.noespansivi}
\end{proposition}
\begin{proof}
  For dimension greater than 1, this is a consequence of the well-known
  result of Shereshevsky \cite{shereshevsky93}. Consider a one-dimensional
  NCA $F$. Suppose that $F$ is expansive with expansive constant
  $\epsilon$. Let $i\in \ene$ be such that $\frac{1}{2^i}<\epsilon$.
  Construct a finite configuration $c_i(j)=1$ if $i=j$, $0$ otherwise.
  Since $F$ is expansive there exists $t_1$ such that 
  $d(F^{t_1}(0),$ $F^{t_1}(c_i))>\epsilon$. For the same reason there
  exists $t_2$ such that $d(F^{t_2}(0),F^{t_2}(c_{-i}))>\epsilon$.
  Therefore there exists a time $t\geq\max\set{t_1,t_2}$ for which
  $F^t(c_i)$ consists of at least two $1$'s, violating $FNC$ condition.
\end{proof}
\begin{lemma}
  \label{lem.contraction}
    Let $F$ be a NCA.
    For any $x\in\ccc_F$, there exists $c_x\in \ene$ such that
    $\forall t\in \ene,\;\len{F^t(x)}\geq c_x$.
\end{lemma}
\begin{proof}
    Let $x\in \ccc_F$. Since each cell can contribute at most $q-1$ to
the sum
    $s(x)$, it holds that
    $\forall t\in \ene,\;\len{F^t(x)}\geq \lceil \frac{s(x)}{q-1}\rceil$.
\end{proof}
The following proposition is a trivial consequence of 
Lemma \ref{lem.contraction}.

\begin{proposition}
   \label{prop.noclass1}
   There are no NCA in class $C_1$.
\end{proposition}

\begin{proposition}
   \label{prop.noclass1cap3}   
   There are no NCA in class $K_1\cap C_3$.
\end{proposition}
\begin{proof}
   Let $F$ be a \ca in $C_3$ and let $c$ be such that
   \begin{equation}
     \label{eq.def.c3}
     \lim_{t\to\infty}\len{F^t(c)}=\infty\enspace .
   \end{equation} 
   We have two cases:
   \begin{enumerate}
     \item $\displaystyle \sup_{t\in \ene} m(F^t(c))=-\infty$; \item
     $\displaystyle \sup_{t\in \ene} m(F^t(c))=c\in \zed$ and $\displaystyle
     \sup_{t\in \ene} M(F^t(c))=\infty\enspace .$
   \end{enumerate} 
   We consider only the former case, the proof for the latter is
   similar.  For any $\delta>0$, let $\bar{n}\in \ene$ be such that
   $\frac{1}{\bar{n}}\leq\delta\leq\frac{1}{\bar{n}-1}$.  By
   \eqref{eq.def.c3}, there exists a time $t$ such that
   $m(F^t(c))=-h>\bar{n}$.  Therefore $d(\uz,\sigma^{-h}(c))\leq\delta$. 
   Since $F$ commutes with $\sigma$, we have
   $$m(F^t(\sigma^{-h}(c)))=m(\sigma^{-h}(F^t(c)))=m(F^t(c))+h=0$$ which
   implies that
   $$d(F^{t}(\sigma^{-h}(\uz)),F^{t}(\sigma^{-h}(c)))=d(\uz,F^{t}
   (\sigma^{-h}(c)))= 1\enspace .$$ Since $\delta$ is arbitrary, last
   inequality implies that $\uz$ is not an equicontinuity point. 
   Since we are in a perfect space we conclude that $F$ is not an
   equicontinuous system.
\end{proof}
A word $W\in S^{\star}$ is called \emph{blocking} word if there exists
an infinite sequence of words $w_n$ such that
\begin{enumerate}
   \item $\forall n\in \ene,\;\len{w_n}=2h+1>r$ with $h\in \ene$; \item
   $\forall c\in S^{Z},\;(c_{-k:k}=W \Rightarrow\forall t\in
   N\setminus\set{0}\; F^t(W)_{-i:i}=w_t)$\enspace .
\end{enumerate}
In other words $W$ partitions the evolution diagram of $F$ in two
completely disconnected parts: perturbations made in one part are
completely ``blocked'' by $W$.  This notion is
fundamental for characterizing \cas dynamics in the presence of
equicontinuity points as we can see in the next proposition.  It will
also be crucial in the proof of Proposition~\ref{prop.c2.k2}.
\begin{proposition}[\cite{blanchardtisseur}]
  \label{prop.bltisseur}
  Any equicontinuity point has an occurrence of a blocking word. 
  Conversely if there exist blocking words, then any point with
  infinitely many occurrences of a blocking word to the left and to
  the right (of $0$) is an equicontinuity point.
\end{proposition}
\begin{proposition}[\cite{blanchardtisseur}]
  \label{prop.bltisseur.dpo}
  Let $X$ be a $\sigma$-transitive SFT. If $F$ has an equicontinuity point
  then ($F$ onto $\Leftrightarrow$ regular).
\end{proposition}

The above proposition implies that surjective NCA in class $K_{2}\cap 
C_{3}$ are regular.

\begin{proposition}[\cite{blanchardtisseur}]
  \label{prop.bltisseur.periodici}
  A \ca $F$ is equi\-continuous and onto if and only if $\exists p>0$
  such that $\forall c\in S^{\zed},\;F^p(c)=c$.
\end{proposition}

A consequence of the above proposition is that surjective NCA in class 
$K_{1}\cap C_{2}$ are periodic and non-surjective ones are ultimately 
periodic.

\begin{proposition}
   \label{prop.c2.k2}
   There are no NCA in class $K_2\cap C_2$.
\end{proposition}
\begin{proof}
   Since we are in a perfect space if a \ca is not equicontinuous
   then there exists at least a configuration $c$ which is not an 
   equicontinuity point. By Proposition \ref{prop.knud} we can assume
   that $c$ is finite. By the hypothesis, there exists a constant $K$
   such that $\forall t\in \ene,\;\len{F^n(c)}\leq K$. Assume that
   $\lim_{t\to\infty}m(F^{t}(c))=-\infty$ (if this is not the case then
   $\lim_{t\to\infty}m(F^{t}(c))=+\infty$ and the rest of the proof
   will be perfectly similar).
   On the other hand, there exists at least an
   equicontinuity point and, by Proposition \ref{prop.bltisseur},
   there exists at least a blocking word $W$. 
   Construct a configuration $d$ such that 
   \[
     \forall i\in\ Z,\;d(i)=
     \left\{
     \begin{array}{ll}
        c(i) &\textrm{if}\; m(c)\leq i\leq M(c)
        \\
        W(i) &\textrm{if}\; 2M(c)\leq i\leq 2M(c)+\len{b}
        \\
        0  &\textrm{otherwise} \enspace .
     \end{array}
     \right.
   \]
   It is clear that $\displaystyle \lim_{t\to\infty}\len{F^{t}(d)}=+\infty$.
\end{proof}

For any \ca $F$, denote $F_F$ the restriction of $F$ to $C_F$ and
$F_P$ the restriction of $F$ to $C_P$. 

\begin{proposition}
  \label{prop.bij-fin}
  A NCA $F$ is surjective if and only if its restriction to $F_F$
  is bijective.
\end{proposition}
\begin{proof}
  One implication is the well-known Moore-Myhill
  theorem~\cite{moore,myhill}.  For the converse, remark that if $F$
  is surjective every finite configuration has a pre-image.  The FNC
  condition grants that this pre-image is finite.
\end{proof}

\noindent
The following definition is essentially taken from~\cite{cattaneo180}.

\begin{definition}
   \label{def.gensubshift}
   A \ca $<S^{\zed},G>$ is said to be a \emph{generalized subshift} on the
   set $U\subseteq S^{\zed}$ iff 
   \begin{enumerate}
      \item
        $U$ is $G$ positively invariant \ie $G(U)\subseteq U$;
      \item
        there exists a mapping $T,T'\colon S^{\zed}\to\ene\setminus\{0\}$
such that
        $\forall c\in U,\; G^{T(c)}(c)=\sigma^{T'(c)}(c)$.
   \end{enumerate}
\end{definition}

The following lemma present peculiar properties of 
the shifting function of generalized subshifts
on $C_F$.

\begin{lemma}
  \label{lem.shiftfunc}
  Let $F$ be a generalized subshift on $C_F$ 
  and let $T,T'$ as in Definition~\ref{def.gensubshift}. 
  Then either $\forall x\in C_F, T(x)=T'(x)$ or 
  $\forall x\in C_F, T(x)=-T'(x)$.
\end{lemma}
\begin{proof}
  By contraposition, suppose that there exist $x,y\in X$ such that
  $T(x)=T'(x)$ and $T(y)=-T'(y)$. Let 
  $n_x=\max_{0\leq i\leq T(x)}{\len{F^i(x)}}$ and
  $n_y=\max_{0\leq i\leq T(y)}{\len{F^i(y)}}$.
  Build the following finite configuration
  \[
    \forall i\in\zed,\;w(i)=
    \left\{
    \begin{array}{ll}
       y(i+\len{y}) &\text{if}\;-\len{y}-n_y\leq i\leq -1-n_y
       \\
       x(i-1) &\text{if}\;1+n_x\leq i\leq \len{x}+n_x
       \\
       0 &\text{otherwise}\enspace .
    \end{array}
    \right.
  \]
  Since $\lim_{t\to\infty}\len{F^t(w)} = +\infty$ there exists no
  $n\in\zed\setminus\set{0}$ such that $F^n(w)=\sigma^n(w)$ or 
  $F^n(w)=\sigma^{-n}(w)$.
\end{proof}
The following lemma links
properties of the evolution of $F$ with set properties of the maps
$F_F$ and $F_P$.
\begin{lemma}
  \label{lem.transbijf}
  Let $F$ be a transitive \ca. Then $F_F$ is bijective and $F_P$ is
  surjective.
\end{lemma}
\begin{proof}
   It is well-known that transitivity implies surjectivity. By
   Moore-Myhill theorem, this last property implies that $F_F$ is
   injective. By Proposition~\ref{prop.knud}, transitive \cas are 
   transitive over $C_F$. This fact implies that $F_F$ is surjective.
   Using Proposition~\ref{prop.knud} again, we have that transitive
   \cas are transitive over $C_P$ and hence $F_P$ is surjective.
\end{proof}
\begin{proposition}
   \label{prop.gsub.chaos}
   Generalized subshift \cas on $\ccc_F$ are 
   transitive and regular. 
\end{proposition}
\begin{proof}
  Let $F$ be a generalized subshift \ca on $\ccc_F$ and $T,T'$ as in
  Definition~\ref{def.gensubshift}.
  We are going to prove that $F$ is transitive on $C_F$; 
  by Proposition~\ref{prop.knud} we have the first part of the thesis.
  For any cylinder set $[w]_t$, let $\bar{w}$ the finite configuration
  such that $\bar{w}_{i+t}=w_{i+t}$ for $0\leq i\leq\len{w}$ and $0$
  elsewhere. Take two arbitrary cylinder sets $[w]_t$ and $[z]_l$.
  Let $n_{\bar{w}}=\max_{0\leq i\leq T(\bar{w})}\len{F^i(w)}$ and
  $n_{\bar{z}}=\max_{0\leq i\leq T(\bar{z})}\len{F^i(z)}$. Let 
  $m=T(\bar{w})T(\bar{z})$. Finally, choose $k,q\in\zed$ such that
  $q>n_{\bar{z}}+n_{\bar{w}}$ and $t=km+q$.
  Build a finite configuration $c$ such that 
  $c=\bar{w}\oplus\sigma^{km+q}(\bar{z})$, where $\oplus$ is the bitwise 
  ``or'' operation.
  It is not difficult to see that $F^{km+q}(c)\in[z]$ and
  $c\in[w]$; this completes the first part of the proof.
  
  Since $C_F$ is dense in $S^{\zed}$, for any configuration 
  $x\in S^{\zed}$, there exist a sequence 
  $\set{u_i}_{i\in\ene}\subseteq C_F$ such that $\lim_{i\to\infty}u_i=x$.
  and let 
  \[ 
     n_{u_i}=\max_{0\leq j\leq T(u_i)}\len{F^j(u_i)}\enspace .
  \]
  For any $u_i$, build a spatial periodic configuration 
  \[
     y_i=0^{2n_{x_i}}x_i0^{2n_{x_i}}0^{2n_{x_i}}x_i0^{2n_{x_i}}\ldots
     0^{2n_{x_i}}x_i0^{2n_{x_i}}\ldots
  \]
  such that $(y_i)_k=(x_i)_k$ for $m(x_i)\leq k\leq M(x_i)$.
  Clearly $y_i$ is ultimately periodic for $F$. By Lemma \ref{lem.transbijf}
  we have that it is in fact periodic. The fact that 
  $\lim_{i\to\infty} y_i = x$ closes the proof.
\end{proof}
\noindent
From Proposition \ref{prop.gsub.chaos} and
\cite{glasnerweiss,kurka97}, generalized subshift \cas on $\ccc_F$ are
sensitive to initial conditions.

When studying asymptotic behavior of a \ca it is of fundamental
importance to understand the structure of the so called
\emph{limit set} $\Omega=\cap_{n\in\ene}F^{n}(S^{\zed})$. This set
can be viewed as the set of configurations which have an infinite
number of predecessors. It is well-known that is compact, non-empty
and that it is the maximal (w.r.t. set inclusion) attracting set.
In the sequel we will very interested in a special subset of the
limit set: $\Omega_F=\cap_{n\in\ene}F^n(\ccc_F)\subseteq\ccc_F$,
that is the set
of finite configurations which have an infinite number of predecessors
(of finite type). One can easily see that for any \ca this set is non-empty
for it should contain at least one homogeneous configuration. 
In the particular case of NCA, the following easy proposition
proves that the cardinality of $\Omega_F$ is
always countably infinite.

\begin{proposition}
  For any NCA, $\Omega$ is an uncountable set while
  the cardinality of $\Omega_F$ is countably infinite.
\end{proposition}
\begin{proof}
  Let $c_1$ be the configuration that has a $1$ in zero and $0$
  elsewhere. Let $h=|m(F^(c_1))|$ (for $a\in\zed$, $|a|$ is the 
  absolute value of $a$). Let $a_b= 0^{2h+1}b0^{2h+1}$ for 
  $b\in\zu$. For any configuration $x\in\zu^{\zed}$, build the 
  configuration $c_x = a_{x_0}a_{x_1}\ldots a_{x_n}\ldots$ Note that
  $c_x\in \Omega$.
  In particular if $x\in C_F\cap\zu^{\zed}$ then $c_x\in\Omega_F$.
  It is easy to see for $x,y\in\zu^{\zed}$, if $x\ne y$ then 
  $c_x\ne c_y$.
\end{proof}

A \ca $F$ is said to be a \emph{ultimately generalized subshift} on
$U\subseteq S^{\zed}$ is $F$ is a generalized subshift on $U$ and
$\forall x\in S^{\zed}$, there exists $n=n_x\in\ene$ such that
$F^{n_x}(x)\in U$. 
\begin{proposition}
 \label{prop.k3c2}
 Non-surjective NCA in $K_3\cap C_2$ are ultimately
 generalized subshifts on $\Omega_F$.  All surjective NCA in class
 $K_3\cap C_2$ are generalized subshifts on $\ccc_F$.
\end{proposition}
\begin{proof}
   Consider a NCA $F$ in $K_3\cap C_2$, we prove that it is a
   generalized subshift on 
   $\Omega_F\subseteq\ccc_F$. 
   Since $F$ is not equicontinuous
   for any finite configuration $c$ either $m(F^t(c))$ or $M(F^t(c))$ is not
   a constant function.  Since $F\in C_2$ there exists $L\in\ene$ such that
   $\forall t\in\ene,\;\len{F^t(c)}\leq L$. Therefore there exist
   $t_1,t_2\leq S^L$ such that $F^{t_1}(F^{t_2}(c))=\sigma^{t_1}(F^{t_2}(c))$.
   Clearly, $\Omega_F$ is positively invariant and, by the above remarks,
   $F$ is a generalized subshift on it, for otherwise one can find
   a finite configuration in which $m(\cdot)$ is constant.
   Moreover, since $\forall t\in\ene,\;\len{F^t(c)}\leq L$, there exists
   $\bar{t}\in\ene$ such  that $F^{\bar{t}}(c)\in\Omega_F$.
   For the second part of the proof remark that surjective NCA are
   bijective on $\ccc_F$ (\cf Proposition \ref{prop.bij-fin}) then
   $\Omega_F=\ccc_F$.
\end{proof}

\begin{proposition}
  \label{prop.ultsubshift}
  Let $F$ be an ultimately generalized subshift on $U\subseteq\ccc_F$
  with $|\Omega_F|=\infty$. Then $F$ is sensitive to initial
  conditions.
\end{proposition}
\begin{proof}
  We prove that they are sensitive to initial conditions on $\ccc_F$,
  by Proposition \ref{prop.knud} we have the thesis.
  
  Consider a finite configuration $c$ and $B_{\delta}$ the ball of
  radius $\delta>0$ centered on $c$. Choose an integer $n$ such that
  $\frac{1}{n}<\delta$. Let $t\in\ene$ be such that $F^t(c)\in\Omega_F$
  (by the hypothesis, $t$ always exists). 
  Let $n_c=\max_{0\leq i\leq T(c)+t} \len{F^i(c)}$. Choose $x\in\Omega_F$.
  By Lemma~\ref{lem.shiftfunc}, without loss of generality we can assume that
  the patterns generated by $c$ and $x$ move to the left.
  Now, choose $h>t$ such that $m(F^h(c))+n_c<0$. 
  Let $n_x=\max_{0\leq i\leq T(x)} \len{F^i(c)}$. Choose $k$ such that
  $kT(x)>h+n_x+n$ and $m(F^{kT(x)}(x))=0$. Finally, build the configuration
  $w = c\oplus \sigma^{kT(x)}(x)$, where $\oplus$ is the usual bit-wise
  ``or'' operation. Is is not difficult to see that 
  $d(c,w)<\delta$ and $d(F^{kT(x)}(c),F^{kT(x)}(w))\geq 1$. 
  Since $c$ and $\delta$ were chosen arbitrarily we conclude that $F$ is
  sensible to initial conditions.
\end{proof}

The following corollary is an immediate consequence of 
Propositions \ref{prop.ultsubshift} and \ref{prop.k3c2}.
\begin{corollary}
  NCA of class $C_2$ that are ultimately subshift on $U\subseteq\ccc_F$ 
  are sensitive to initial conditions.
\end{corollary}
\subsection{Examples}\label{examples}
This section proposes several examples which illustrate representative
behaviors in each class.
\begin{example}[Class $K_1\cap C_2$ is not empty]
   \label{ex.id}\mbox{}\\
  $K_1\cap C_2$ contains the identity \ca.$\square$
\end{example}
\begin{example}[Class $K_3\cap C_2$ is not empty]
   \label{ex.184}\mbox{}\\
   Consider the NCA with the following local rule $f$ on the alphabet
   $S=\set{0,1}$:
   \begin{center}
   \begin{tabular}{c|c|c|c|c|c|c|c|c}
            $(x,y,z)$ & $000$ & $001$ & $010$ & $011$ & $100$ & $101$ &
$110$ & $111$ 
     \\
     \hline $f(x,y,z)$&  $0$   & $0$  &  $0$  &  $1$  &  $1$  &  $1$  & 
$0$  &  $1$
   \end{tabular}     
   \end{center}
   Rule $f$ is called ``rule $184$'' in \cite{boccara} and it often
given an
   elementary example of car traffic modeling: symbols $1$'s are cars
moving on a
   highway ($0$'s). When a $1$ finds a $0$ on its right (that a free
path on the
   highway) it moves one cell to the right. If the cell is not free then
it waits.
   $\square$
\end{example}
\begin{example}[Class $K_2\cap C_3$ is not empty]
   \label{ex.k2c3}\mbox{}\\
   Consider the following local rule $f$ on the alphabet $S=\set{0,1,2}$:
   \begin{center}
   \begin{tabular}{l|c|l|l|l|l}
     $(x)$        & $2a$ & $1b$ & $12$ & $0a$
     \\
     \hline $f(x)$& $2$ & $b$  & $1$  & $0$
   \end{tabular}
   \end{center}
   where $a\in A$ and $b\in\set{0,1}$.  Let $F$ be the global rule
   induced by $f$.  It is clear that $W=2$ is a blocking word.  Define
   three configurations $c,d,e$ respectively as $c(0)=2,\,c(1)=1,\,0$
   otherwise; $d(0)=1,\,d(1)=2,\,0$ otherwise; and
   $e(-1)=1,\,c(0)=0,\,c(1)=2,\,0$ otherwise.  Since $\displaystyle
   \sup_{t\in \ene}\len{F^t(c)}=\infty$ and  $F^{-1}(d)=\set{d,e}$,
   by Moore-Myhill theorem~\cite{moore,myhill}, $F$ is not surjective
   and hence not regular. 
   Finally, since $\uz$ is not an equicontinuity point we have that
   $F\in K_2\cap C_3$. $\square$
\end{example}

Cellular automata in class $K_3\cap C_3$ seems to have a more rich variety of
behaviors than other classes; as a consequence is much more difficult
to have clear-cut situations: we can have system of particles with
delayed propagation (Example~\ref{ex.k3c3}) and others in which particles
cross each other without interaction (Example~\ref{ex.no-interaction}).
We also underline that it is exactly in this class that reversible
NCA capable of universal computation have been found \cite{morita98,morita99}.

\begin{example}[Class $K_3\cap C_3$ is not empty]
   \label{ex.k3c3}\mbox{}\\
   Consider the NCA with the following local rule $f$ on the alphabet
   $S=\set{0,1,2}$:
   \begin{center}
   \begin{tabular}{c|c|c|c|c|c|c}
            $(x,y,z)$ & $00b$ & $002$ & $01a$ & $02a$ & $10a$ & $11a$ 
     \\
     \hline $f(x,y,z)$& $0$   & $2$   & $a$   & $0$   & $1$   & $a$  
   \end{tabular}     
   \end{center}
and
   \begin{center}
   \begin{tabular}{c|c|c|c|c|c}
            $(x,y,z)$ & $12a$ & $20b$ & $202$ & $21a$ & $22a$
     \\
     \hline $f(x,y,z)$& $1$   & $0$   & $2$   & $a$   & $2$
   \end{tabular}     
   \end{center}
   One can  verify that for any finite configuration $c$ there 
   exists a time $\bar{t}$ such that for any $t\geq\bar{t}$, $F^t(c)$ 
   is such that all symbols $1$
   (if any) are on the positive coordinates, while symbols of type $2$
   are on the negative part. $1$'s and $2$'s propagate in opposite directions.
   Normally type $2$ symbols propagate at speed $1$ to the left
   (one unit displacement per time unit). This speed reduces of one half if
   $2$ encounters a $1$ like in $102$; in a sense type $1$'s symbols have
   precedence over $2$'s. Therefore for any $\epsilon>0$ choose
   $\bar{n}\in \ene$ such that 
   $\frac{1}{\bar{n}}\leq\epsilon\leq\frac{1}{\bar{n}-1}$. 
   Construct the following configuration $\bar{c}$ as follows:
   \[
      \forall i\in \zed, \bar{c}(i)=
      \left\{
      \begin{array}{ll}
         c(i) &\mbox{if}\;m(c)\leq i\leq M(c)
         \\
         2    &i=\max\set{\bar{n},\bar{t}}
         \\
         0    &\mbox{otherwise}\enspace .
      \end{array}
      \right.
   \]
   Then there exists $h\leq 2\max\set{\bar{n},\bar{t}}$ such that
   $d(F^h(c),F^h(\bar{c}))\break =1$. From the fact that 
   $d(c,\bar{c})\leq\epsilon$ and Proposition~\ref{prop.knud}
   we have that $F$ is sensitive to initial
   conditions. Note that $F$ is not surjective since
   its restriction to configurations on the alphabet $\{0,1\}$ coincides with
   elementary rule $184$ of Example~\ref{ex.184} which is not surjective.
   $\square$
\end{example}

\begin{example}
  \label{ex.no-interaction}
  Consider a system which describes the interaction of two
  particles (denoted by 1 and 2) in a neutral media (symbols 0).
  Particles of type 1 move left to right while type 2
  move in the opposite direction. A symbol 3 denotes the presence
  of a particle of type 2 and one of type 1 on the same site.
  These elementary rules are formalized by the following local
  rule:
   \begin{center}
   \begin{tabular}{c|c|c|c|c}
            $(x,y,z)$ & $0xa$ & $1xa$ & $2xa$ & $3xa$
     \\
     \hline $f(x,y,z)$&  $0$   & $1$   & $0$   & $1$ 
   \end{tabular}
  \end{center}     
  and
  \begin{center}
   \begin{tabular}{c|c|c|c|c}
            $(x,y,z)$ & $0xb$ & $1xb$ & $2xb$ & $3xb$
     \\
     \hline $f(x,y,z)$&  $2$   & $3$   & $2$   & $3$ 
   \end{tabular}     
   \end{center}     
  where $a\in\set{0,1}$, $b\in\set{2,3}$ and $x\in\set{1,2,3,4}$.

  It is easy to see that $F$ is injective and hence
  its restriction to spatial periodic configuration is
  injective (see \cite{durfiesta96}). The last property
  implies that $F$ is regular. 
  $F$ is also transitive. In fact, for any finite configuration
  $x=x_{-k}\ldots x_{-k+h}$ ($k,h\in N$) construct a configuration
  $y$ such that for $-k-h\leq i< -k$, $y(i)=1$ or $3$ if $x(i+h)=1$, 
  $y(i)=0$ otherwise. Moreover for $-k+h< i\leq -k+2h$ let
  $y(i)=2$ if $x(i-h)=2$ or $3$. Remark that $F^h(y)=x$. For any 
  pair of cylinders $C_1=C(w_{-k},\ldots,w_{-k+h})$ and
  $C_2=C(z_{-g},\ldots,z_{-g+n})$ ($k,h,g,n\in N$). Choose $c\in C_1$.
  Let $t$ be such that $m(F^t(c))<-g+h$ and $M(F^t(c))>-g+n+h$. 
  Define $d$ such that for $-g\leq i\leq -g+n$, 
  $d(i)=z_i$. Moreover, for $m(F^t(c))\leq i\leq m(F^t(c))+h$ let $d(i)=1$ if
  $w_{i-m(F^t(c))}=1$ or $3$; for $M(F^t(c))-h\leq i\leq M(F^t(c))$ let
  $d(i)=2$ if $w_{i-M(F^t(c))}=2$ or $3$. Finally, $d(i)=0$ in all other cases.
  By construction $d\in C_2$. 
  Using previous remark one can see that $F^t(d)(j)=w_j$ for $-k\leq j\leq -k+h$,
  that is to say that $F^t(d)\in C_1$.
  $\square$
\end{example}

Table~\ref{bureau} summarizes the situation.
\begin{table}[htb]
  \begin{center}
    \begin{tabular}{l|c|c|c|}
       &$C_1$&$C_2$&$C_3$
       \\
       \hline $K_1$&$\emptyset$ (Prop.~\ref{prop.noclass1})
                   &Example \ref{ex.id}
                   &$\emptyset$ (Prop.~\ref{prop.noclass1cap3})
       \\
       \hline $K_2$&$\emptyset$ (Prop.~\ref{prop.noclass1})
                    &$\emptyset$ (Prop.~\ref{prop.c2.k2})
                    & Example \ref{ex.k2c3}
       \\
       \hline $K_3$&$\emptyset$ (Prop.~\ref{prop.noclass1})
                    &Example \ref{ex.184} & Example \ref{ex.k3c3}
       \\
       \hline $K_4$&$\emptyset$ (Prop.~\ref{prop.noespansivi})
                   &$\emptyset$ (Prop.~\ref{prop.noespansivi})
                   &$\emptyset$ (Prop.~\ref{prop.noespansivi})
    \end{tabular}
  \end{center}
  \caption{NCA in K\r{u}rka's and Cattaneo's classifications.}
  \label{bureau}
\end{table}
\section{Conclusions and open problems}
The paper discusses three formulations of the notion of being 
number-con\-ser\-ving which have arisen in recent works on the study of
physical phenomena ruled by conservation laws of mass or energy.
We have proved that all these formulations are equivalent and we
have given another equivalent definition in terms of density.

Moreover, we have given a linear time algorithm for deciding if a
CA (of any dimension and any state set) is number conserving.
We stress that the linearity of the algorithm is something
rather surprising since most of the interesting structural or
dynamical properties of CA are known to be undecidable.

The strong constraints imposed by number-conservation simplifies the
study of asymptotic evolutions. Results can be conveniently 
interpreted in terms of particles system. Class $K_1\cap C_3$
is composed by systems with ``non-moving'' particles. In class
$K_3\cap C_2$ we find systems in which particles move along only one
direction with identical speed (of course speed may vary from system
to system). Systems in class $K_2\cap C_3$ allow particles in 
many different directions (possibly opposite) with the peculiarity that
particles have blocking collisions. Finally, systems in class
$K_3\cap C_3$ have the same properties as those in class $K_2\cap C_3$
but with non-blocking collisions.

Although many aspects have been understood there are still many
interesting open problems that we think they worth further study.
For example, we have already said that surjectivity property is
undecidable for CA in general. But maybe it turns out to be
decidable for the sub-class of NCA. Surjectivity plays a prominent
role in the study of CA dynamics since it is a necessary condition
(for example) for many popular definitions of chaotic dynamics.
In Section~\ref{subsec.dynamics} we have found that, with exception for class 
$K_3\cap C_3$, surjectivity is equivalent to the property of
having a dense set of periodic points (again, this last property
is one of the component of Devaney's definition of chaos).
For class $K_3\cap C_3$ we do not know if this property holds or not.
\bibliographystyle{abbrv}
\bibliography{enrico,ours}
\end{document}